\documentclass[twocolumn,preprint]{aastex62}
\usepackage[T1]{fontenc}
\usepackage[english]{babel}
\usepackage{natbib}
\usepackage{ulem,xcolor}
\usepackage{amsmath, graphicx}
\usepackage{soul}
\def\rhessi{{\textit{RHESSI}}}

\def\sdo{{\textit{SDO}}}
\def\mw{{microwave}}

\definecolor{dark green}{rgb}{0.0, 0.5, 0.0}

\begin{document}

\title{Coronal Heating Law Constrained by Microwave Gyroresonant Emission}

\author[0000-0001-5557-2100]{Gregory D. Fleishman}
	\affil{Center For Solar-Terrestrial Research, New Jersey Institute of Technology, Newark, NJ 07102, USA}

\author[0000-0002-1107-7420]{Sergey A. Anfinogentov}
\affil{Institute of Solar-Terrestrial Physics, Irkutsk 664033, Russia}

\author[0000-0002-5453-2307]{Alexey G. Stupishin}
\affiliation{Saint Petersburg State University, 7/9 Universitetskaya nab., St. Petersburg, 199034 Russia}


\author[0000-0001-8644-8372]{Alexey A. Kuznetsov}
\affiliation{Institute of Solar-Terrestrial Physics, Irkutsk 664033, Russia}

\author[0000-0003-2846-2453]{Gelu M. Nita}
	\affil{Center For Solar-Terrestrial Research, New Jersey Institute of Technology, Newark, NJ 07102, USA}

\begin{abstract}
The question why the solar corona is much hotter than the {visible} solar surface still puzzles solar researchers. Most theories of the coronal heating involve a tight coupling between the coronal magnetic field and the associated thermal structure. This coupling is based on two facts: (i) the magnetic field is the main source of the energy in the corona and (ii) the heat transfer preferentially happens along the magnetic field, while is suppressed across it. However, most of the information about the coronal heating is derived from analysis of EUV or soft X-ray emissions, which are not explicitly sensitive to the magnetic field. This paper employs another electromagnetic channel---the {sunspot-associated} \mw\ gyroresonant emission, which is explicitly sensitive to both the magnetic field and thermal plasma. We use nonlinear force-free field reconstructions of the magnetic skeleton dressed with a thermal structure as prescribed by a field-aligned hydrodynamics to constrain the coronal heating model. We demonstrate that the \mw\ gyroresonant emission is extraordinarily sensitive to details of the coronal heating. We infer  heating model parameters consistent with  observations.
\end{abstract}

	





\section{Introduction}

Why the outer region of solar atmosphere, the solar corona, is much hotter than the visible surface of the Sun, the photosphere, remains one of the greatest challenges in solar and stellar physics. Comparison of the full Sun images in the extreme ultraviolet (EUV) or soft X-ray (SXR) ranges with photospheric magnetograms shows that the EUV/SXR brightness from the areas with strong magnetic field, active regions (ARs), is noticeably larger than from the quiet Sun areas. This implies that release of the magnetic energy plays a role in the coronal plasma heating \citep[e.g.,][and references therein]{Klimchuk2006,Klimchuk2015}. However, neither the exact physical mechanism for the coronal heating nor phenomenological relationship  between the magnetic field properties (e.g., strength, twist, etc.) and thermal coronal plasma have yet been established. Several physical processes have been proposed for the coronal heating \citep[see an overview by][]{Mandrini2000}, which fall in two large groups: stressing models and wave models. Each  model predicts its unique scaling between such properties of a magnetic flux tube as its length, $L$, and the mean (or a footpoint) magnetic field $B_{avg}$ (or $B_{foot}$) and the resulting volumetric heating rate $Q$, often parametrized as power-laws $Q\propto B_{avg}^a/L^b$.  Once these properties have been obtained over a representative subset of coronal magnetic flux tubes, the scalings could be straightforwardly derived from the data. The challenge is that it is extremely difficult to isolate and reliably measure properties of individual coronal flux tubes.

Routinely available approaches to probing thermal coronal plasma rely on either EUV or/and SXR data. These emissions in the corona are optically thin, which has both advantages and disadvantages. The advantage is that the entire corona can be probed at once as the EUV/SXR intensity (from a given line of sight, LOS) is determined by distribution of the thermal plasma along the entire LOS. This includes distributions of density, temperature, and elemental abundances. By the same token, this makes extremely difficult to isolate any localized contribution to the emission, e.g., from a flux tube of interest. In addition, elemental abundances can vary in both space and time. Finally, neither EUV nor SXR emission depend on the magnetic field explicitly; thus, it is difficult to link the thermal properties with the underlying magnetic properties. Numerous attempts of performing such analysis with EUV or SXR data have yielded somewhat controversial results \citep{2004ApJ...615..512S, 2006ApJ...645..711W,2007ApJ...666.1245W,2008ApJ...689.1388L,2008ApJ...676..672W,2011A&A...531A.115D, 2017ApJ...846..165U, 2019ApJ...877..129U,2019AGUFMSH41F3323S} overall favoring the ranges 0.2--1 for both $a$ and $b$ \citep{2019ApJ...877..129U}.

A complementary, but largely unexplored approach, is the use of coronal thermal radio emission. Main advantages of the radio emission in addressing the coronal heating problem are (i) the radio emission is explicitly sensitive to the magnetic field; (ii) it can be both optically thin and thick; (iii) the magnetic-field-sensitive gyro emission is unique as it is typically optically thick from several small gyroharmonics ($s=$2, 3) and, in addition, does not depend on the elemental abundances. To fully utilize the potential of the microwave emission in addressing the coronal heating problem, spatially resolved data at many frequencies would be desirable \citep{2020AAS...23538501G}.

In this paper we employ microwave data at only two frequencies focusing on gyro emission at 17\,GHz. We selected an AR with a favorable range of the photospheric magnetic field values that produces optically thick and thin gyro emission at 17\,GHz in multiple locations. We employ three-dimensional (3D) modeling with GX Simulator \citep{Nita2018} based on several reconstructions of the coronal magnetic field \citep{2017ApJ...839...30F} which we fill with a thermal plasma according to a range of parametric heating models. From each 3D magneto-thermal model, we compute synthetic radio maps and compare them with observations. The best model, which offers an almost perfect match with the  observational data  employed in this study, favors a parametric heating with $a=1$ and $b=0.75$.


\section{Methods}
\subsection{Gyroresonant Mechanism of Microwave Emission}

Radio emission from the quiescent corona is formed by two processes---free-free and gyro emissions \citep{1962SvA.....6....3Z, 1962ApJ...136..975K}.  The gyro emission from thermal plasma with $T\lesssim10$\,MK is commonly called gyroresonant (GR) emission as it typically has large optical depth at a given frequency $f$ at only several resonant layers where the frequency matches small integer multiples ($s=1,\, 2,\, 3...$) of the gyro frequency $f_B$,
\begin{equation}
\label{Eq_f_gyro}
f {\rm [Hz]} = sf_B = 2.8\times10^6 sB{\rm [G]}.
\end{equation}
For typical coronal temperatures, $T\sim 1-3$\,MK, $s=3$ is the largest optically thick gyro harmonics \citep[e.g.,][]{2007SSRv..133...73L}. The brightness temperature $T_B$ of thermal optically thick emission is equivalent to the plasma temperature $T$, which allows the conversion $T_B(f)$ to $T(B)$
\citep{Nita_etal_2011, 2011ApJ...728....1T, 2015ApJ...805...93W}. This means that gyroresonant microwave emission provides spatially resolved information on the coronal magnetic field and associated thermal structure simultaneously:
the morphology (shapes) of the microwave emission at various frequencies yields the spatial distribution of the coronal magnetic field, while the brightness of this emission directly yields the electron temperature of the coronal plasma.

This simple picture of optically thick GR emission from the third gyro harmonics has important limitations that, in fact,  help model validation. The GR opacity depends on the viewing angle between the LOS  and the direction of the magnetic field vector. The opacity is small within a conical region along the magnetic field, which results in a longitudinal transparency window. The angle of this cone, which is typically within 10--20$^\circ$, depends on $T$, $n_e$, $s$, and the wave-mode $\sigma$: $\sigma=1$ for the ordinary (O) mode and $\sigma=-1$ for the extraordinary (X) mode. This means that the GR emission from the third gyro layer can be weak if  it is observed within such a transparency window. Then, for higher temperatures, $T\gtrsim4$\,MK, {gyro-resonant layers at} higher harmonics ($s=4,\, 5$) can {have a significant optical thickness to} 
produce unexpectedly bright gyro emission from regions with relatively small magnetic field. 

\begin{figure}[!h]
\includegraphics[width=1\columnwidth]{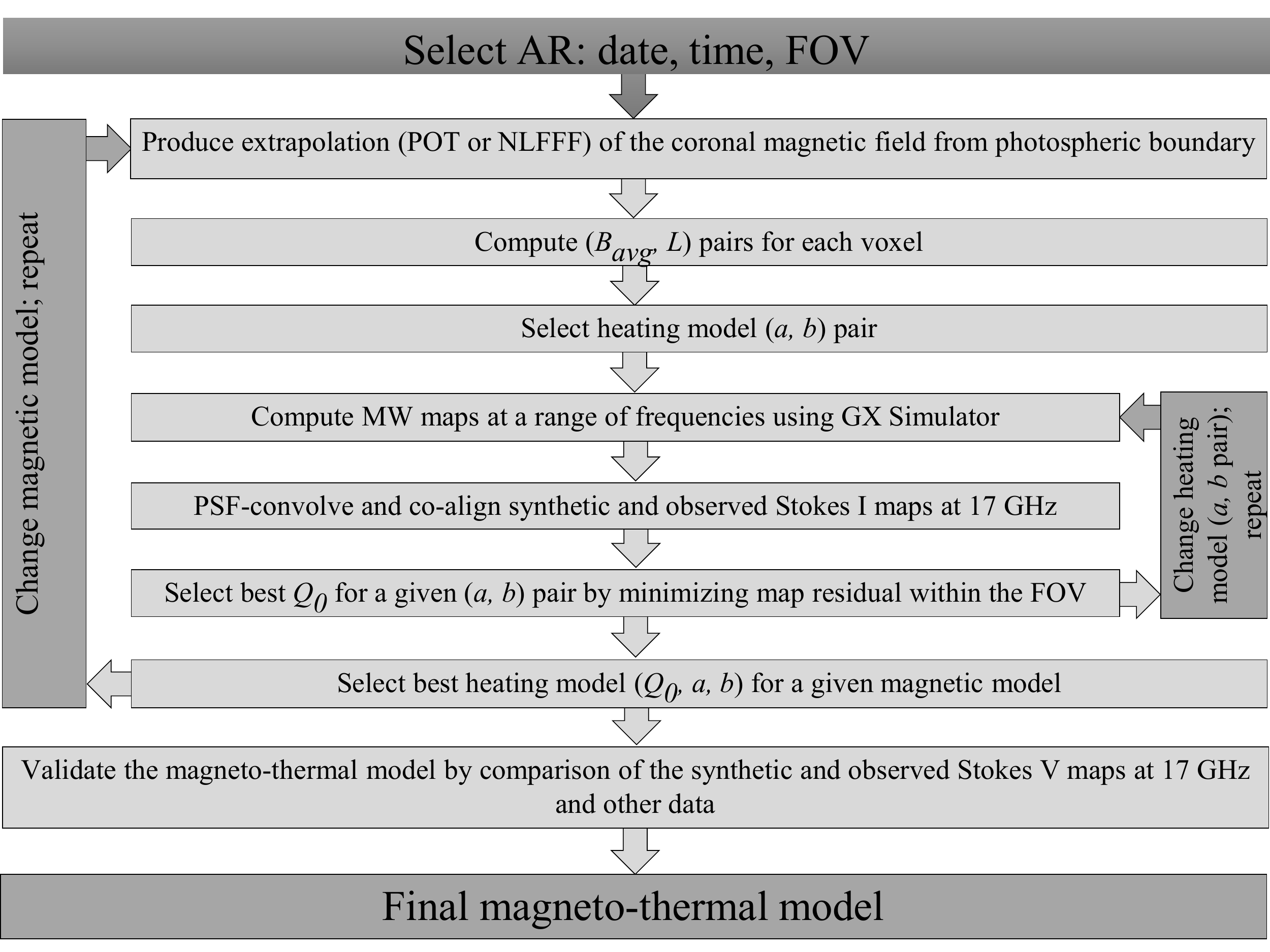}
\caption{Workflow showing main steps of production and validation of the magneto-thermal models of an AR. Input, outcome, and iterative steps are given on top of dark gray background.
\label{Fig_Work_flow}
}
\end{figure}

This implies that the microwave GR emission is sensitive to details of  the spatial distribution of  coronal magnetic field and thermal plasma. We employ this sensitivity to constrain a likely parametric heating model of the solar corona.  Below, we devise and validate  a set of 3D magneto-thermal models, as illustrated in the work flow chart shown in Figure\,\ref{Fig_Work_flow} and described in the remainder of this section.

\subsection{Selection of Active Region---AR 11520}

In this study, we employ optical/magnetic field data available from the
\textit{Solar Dynamics Observatory}/Helioseismic and Magnetic Imager \citep[\sdo/HMI,][]{2012SoPh..275..207S}  and \mw\ data available from the Nobeyama Radioheliograph \citep[NoRH,][]{1994IEEEP..82..705N} at 17\,GHz and the Siberian Solar Radio Telescope \citep[SSRT,][]{2003SoPh..216..239G} at 5.7\,GHz. Our focus is on  the 17\,GHz data where we searched for reasonably complex cases with multiple bright sources, which might be produced by a combination of optically thick and thin GR emission. Among several candidates, we selected the AR\,11520 while around the solar disk center, which later produced one of the largest solar eruptions ever observed \citep{2013SpWea..11..585B}.

\begin{figure}
\centering
	\includegraphics[width=0.98\linewidth]{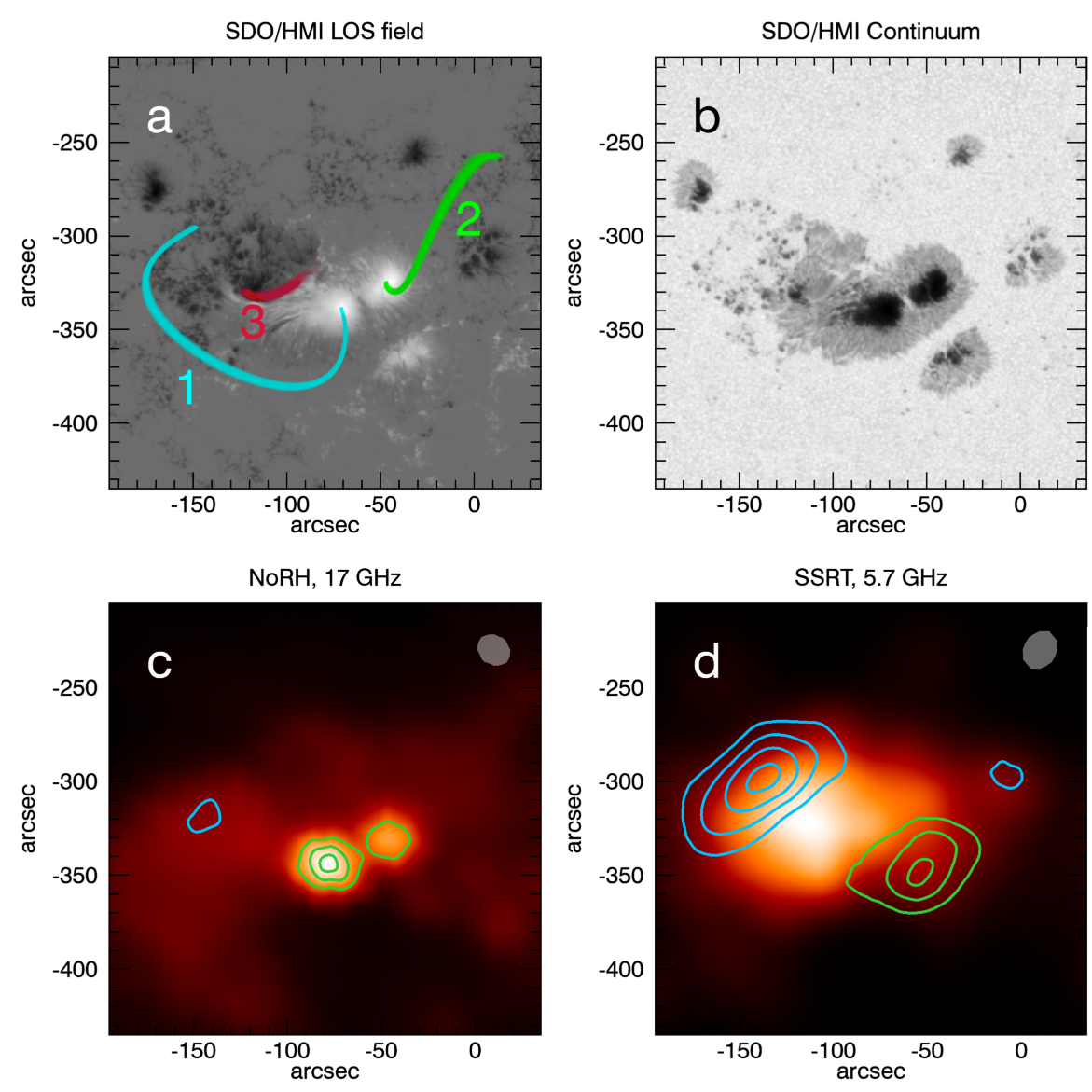}
		\caption{Images of AR NOAA 11520 taken at 2012-Jul-12 05:00\,UT: \sdo/HMI LOS magnetic field (a), \sdo/HMI Continuum (b), NoRH image at 17\,GHz at 05:10\,UT; the peak brightness temperature is 360.6\,kK  (c), and SSRT  daily averaged observations at 5.7\,GHz; the peak brightness temperature is 2,670.9\,kK (d). Microwave intensity images (c, d) are over-plotted  with contours showing positive (green) and negative (blue)   polarized brightness  of the radio emission.
		{ Contours correspond to the Stokes V brightness at the levels of [ -4, 75, 150, 225, 300, 375, and 450]\,kK for NoRH and [-600, -500, -400, -300, -200, -100, 50, 100, 150, 200, 250, and 300]\,kK for SSRT data. {The filled gray ellipses in the upper right corners of panels (c) and (d) show the FWHM instrumental beams for NoRH and SSRT data. The three numbered colored bundles in panel (a) are flux-tubes demonstrating the NLFFF magnetic field connectivity of three regions (umbrae)  with the strongest magnetic field, see Table \ref{BLtable}. They are color coded to facilitate further analysis and discussion.}}  \label{fig_overview}}
	\end{figure}

Figure\,\ref{fig_overview} shows an overview of the data used in our analysis, on  2012-Jul-12, around 05:00\,UT. Panel (a) shows the photospheric LOS magnetogram; the corresponding vector magnetic data are used as a bottom boundary condition for coronal magnetic reconstructions using either potential (POT) or our validated nonlinear force-free field (NLFFF) extrapolation codes \citep{2017ApJ...839...30F}; three sets of field lines obtained from the best magnetic model are color coded to facilitate our analysis. {Flux tubes 1 and 2 originate from centers of two large umbrae with positive magnetic polarity, while flux tube 3 originates from a smaller umbra with negative magnetic polarity located next to a neutral line. This flux tube projects very close to the neutral line}. Panel (b) shows the photospheric  white light map, which is used to create a model of solar chromosphere \citep{Nita2018}. {This chromospheric model does not have any free parameter to be fine tuned in our modeling; but it is needed to define the location-dependent height of the transition region (TR), where the hot corona begins.} Panel (c) shows the  brightness distribution of \mw\ emission at 17\,GHz along with the polarization information. This is the main data source employed to fine tune and validate the 3D magneto-thermal models of the AR. Panel (d) shows  the \mw\ brightness distribution at a lower  frequency, 5.7\,GHz,  which is used to double check the model validity in   the 3D domain. The brightness peak of the 5.7\,GHz emission projects onto a neutral line {and flux tube 3}, although in our case the brightness temperature of this source is not  as high as the ``peculiar'' neutral-line-associated sources reported by \citet{1977ApJ...213..278K,2008SoPh..249..315U} and others.

Figure\,\ref{fig_overview}c displays two bright, right-hand circularly polarized (RCP; green contours), GR sources that project onto two largest umbrae in the center of panel (b), which are associated with two patches of strong positive magnetic field in panel (a). A weaker \mw\ source, extending towards a left-hand circularly polarized (LCP; blue contours) source, projects on a region of a negative magnetic field and a neutral line in panel (a). 

Eqn\,(\ref{Eq_f_gyro}) implies that having a gyro resonance with $s=3$ at 17\,GHz requires $B_{res}\approx 2024$\,G. Inspection of the photospheric magnetogram  in these three regions shows that the maximum magnetic fields are $\approx$ 3280, 3490, and 2600, respectively, which exceed the resonant value $B_{res}$  in all these  regions. Thus, the LOS towards each of these regions must intersect the resonant level at a certain height. If this happens at a coronal height, where the plasma is hot, the GR source  should be bright; if at a chromospheric height, where the plasma is cool, the GR source  should be dim. From our 3D reconstructions described below, the estimated maximum values at the 1\,Mm height are $\approx$ 2620, 2830, and 1880\,G, respectively. This is consistent with having only two bright GR sources at 17\,GHz; the third one would also be bright if the fourth gyro harmonics were optically thick, which is not observed. This offers very stringent constraints on the coronal temperature distribution and makes this AR highly suitable for studying the coronal heating.

\subsection{3D Reconstructions of the Coronal Magnetic Field}
\label{S_POT_NLFFF}

In this study we employ a few different methods of coronal magnetic field reconstruction. Perhaps, the simplest one is the potential field reconstruction (POT) which uses a vertical $B_z$ component of the photospheric magnetogram. We use a very fast (although, not necessarily the most precise) fast Fourier transform (FFT) implementation of the POT reconstruction, which is an integral step of the automated model production pipeline \citep[AMPP,][Nita et al. 2020 in prep.]{Nita2018b}  in  the GX Simulator  package \citep{Nita2018}. In addition, we use several NLFFF reconstructions that utilize full vector boundary condition based on: (i) the AS code  that follows the weighted optimization algorithm \citep{2004SoPh..219...87W}, which is also part of the AMPP; (ii) the IM code that follows full optimization algorithm \citep{2000ApJ...540.1150W} with varying boundaries; and (iii) the IM0 code---full optimization algorithm with fixed boundaries. The AS and IM codes were described and validated by \citet{2017ApJ...839...30F}.






\subsection{Modelling thermal structure}
		
\begin{figure}
    \centering
\includegraphics[width=0.98\textwidth]{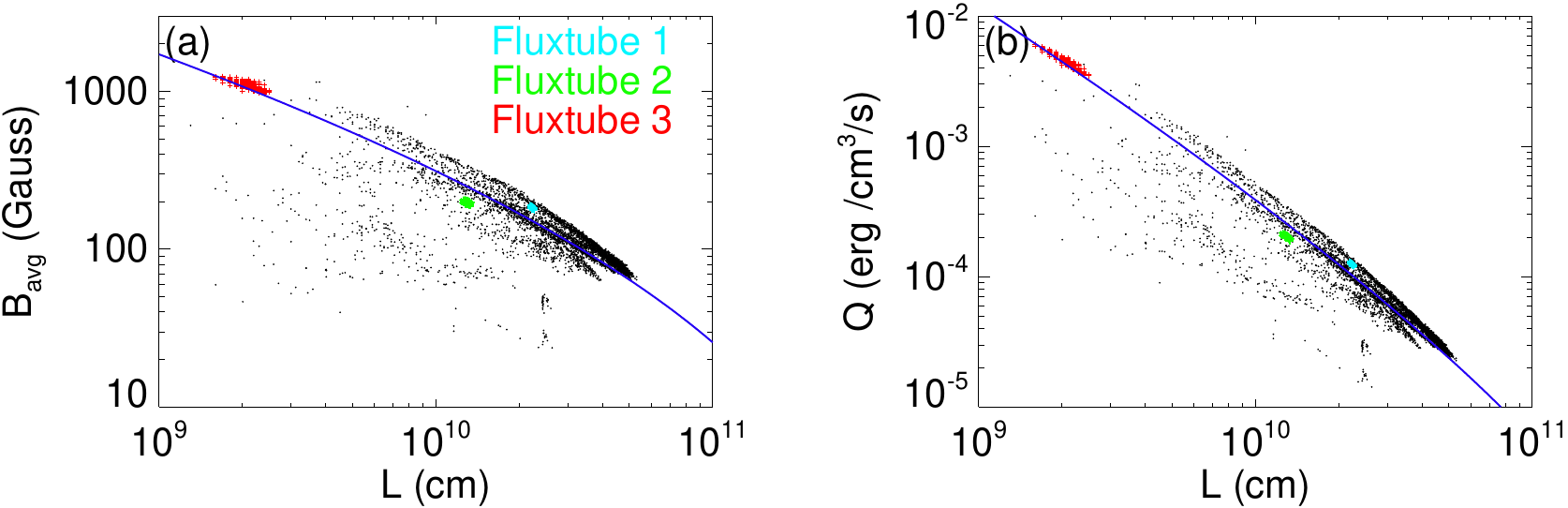}
    \caption{Averaged magnetic field (panel a) and ($a=1$, $b=0.75$) heating rate (panel b) versus magnetic field line length, for all volume voxels (black symbols) and the three color-coded fluxtubes shown in Figure\,\ref{fig_overview}a. 
    The blue lines over-plotted on each panel represent an empirical fit that assumed an empirical-inferred non-linear dependence given by $\log^2 B_{avg}=48.61     -1.84\log{L}$.}
    \label{fig:BQL}
\end{figure}

		
\cite{Nita2018} described the general approach to dressing the magnetic skeleton with a thermal plasma within the GX Simulator modeling framework. It is 
based on 
the field-aligned hydrodynamic models with assumed  heating on the individual magnetic flux tubes defined by the extrapolated field line structure. The volumetric heating rate on a given field line is modelled as a power-law of the magnetic field $B_{avg}$ averaged along a field line and its length $L$

\begin{equation}\label{eq:QvsBL}
Q = Q_0  \left(\frac{B_{avg}}{B_0}\right)^a \left(\frac{L_0}{L}\right)^b,\quad {\rm [erg~cm^{-3}~s^{-1}]}
\end{equation}
where $a$ and $b$ are power-law indices depending on the heating mechanism, $B_0=100$\,G is a typical (normalization) magnetic field, $L_0=10^9$\,cm$^{-3}$ is the normalization length, $Q_0$ is a typical heating rate. This form of the volumetric heating rate is often referred to as parametric heating law. To apply such a heating law to every voxel associated with a closed magnetic field line, GX Simulator processes the 3D magnetic data cube to create all relevant field lines and compute their $B_{avg}$ and $L$. Figure\,\ref{fig:BQL} displays the scatter plot of $B_{avg}$ and $L$ values for the best magnetic reconstruction obtained in our study in panel (a) and a corresponding scatter plot of $Q$ and $L$ values obtained for the same model dressed with an optimal parametric heating law in panel (b).

Once these properties have been computed and the heating law selected, GX Simulator applies precomputed lookup tables  to fill each voxel with the corresponding differential emission measure (DEM) distribution. These lookup tables are obtained from  hydrodynamic simulation code---Enthalpy-Based Thermal Evolution of Loops  \citep[EBTEL,][]{Klimchuk2008, Cargill2012, Cargill2012a, Bradshaw2016, 2017ApJ...846..165U}. GX Simulator contains two lookup tables, described by \citet{Nita2018}; see their Eqs.\,(4),\,(5) and the associated text. One of them,  EBTEL-steady, assumes small, but very frequent heating episodes, which results is essentially steady heating. The other one, EBTEL-impulsive, 
assumes more seldom, but proportionally stronger heating episodes (having a symmetric triangular shape with duration of 20\,s, repeating every 10,000\,s) with long cooling phase---the case of the impulsive heating. 
The default lookup table, used in this study, is EBTEL-impulsive, as the impulsive heating is favored by observations 
\citep[e.g.,][]{Klimchuk2015}. 

The release of GX Simulator described by \citet{Nita2018} employed a rather time consuming method of field line computation and a time consuming interpolation of the DEM from the neighboring $B_{avg}$ and $L$ grid points of the EBTEL lookup tables. Since then, the tool was substantially upgraded such as the speed of the field line computation increased greatly and a few faster approaches to the DEM interpolation were developed. In this study we employ the fastest ``nearest index'' approach, which greatly facilitates producing a large number of thermal models needed in our analysis. Our tests show that this method slightly underestimate the heating for a given $Q_0$, which can easily be compensated by a slight increase of $Q_0$. To compute \mw\ emission we employed moments of the DEM distributions:

\begin{eqnarray}
\label{Eq:n-T}
   n_{e}&=&\left(\int\mathrm{DEM}(T)\,\mathrm{d}T\right)^{1/2},\\\nonumber
   T&=&\frac{\int T\cdot\mathrm{DEM}(T)\,\mathrm{d}T}{n_{e}^2}.
\end{eqnarray}



\subsection{Sensitivity of the Gyroresonant Microwave Emission to the Coronal Parametric Heating}

The plasma temperature in a given flux tube is determined by a volumetric heating rate $Q$ and the length of the loop $L$. It can be approximated by a power-law
\begin{equation}\label{eq:TvsQL}
T \propto  Q^\alpha L^\beta,
\end{equation}
with indices $\alpha$ and $\beta$ varying depending on the heating details. For example, $\alpha=2/7\approx0.286$ and $\beta=4/7\approx0.57$ for a classical steady heating \citep{1978ApJ...220..643R}; see also  \citep{2008ApJ...689.1406P, Nita2018}. 

\begin{figure*}
    \centering
\includegraphics[width=0.9\textwidth]{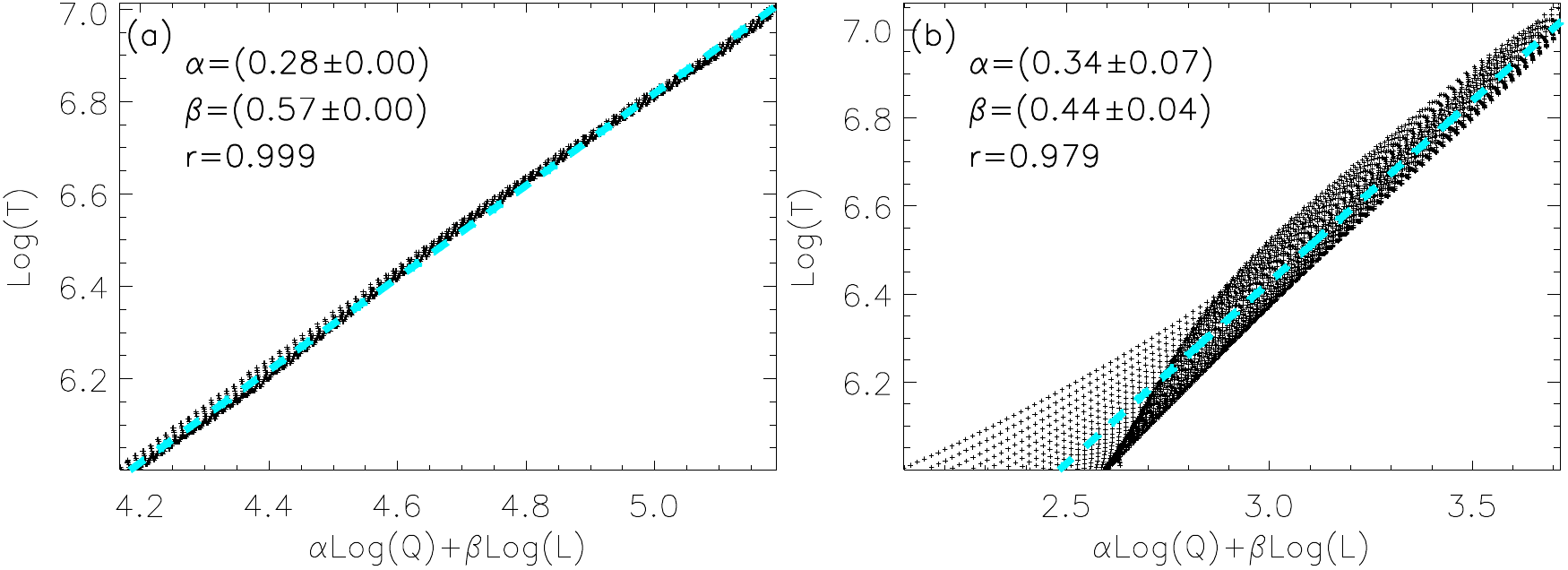}
    \caption{Scatter plots and best linear regression laws obtained with the cross-correlation for two EBTEL lookup tables available at GX Simulator---the steady state (a) and impulsive (b). The indices $\alpha$ and $\beta$ of the regression laws are given in the panels along with the correlation coefficient $r$.
    \label{Fig_AR11520_DEM}  }
\end{figure*}

We computed the mean temperatures using Eq.\,(\ref{Eq:n-T}) for all gridpoints for both lookup tables separately, and then determined the best indices from fitting on the clouds of the data points.
Figure \ref{Fig_AR11520_DEM}  shows the results of this fitting for EBTEL-steady (panel a) and EBTEL-impulsive (panel b) tables. For EBTEL-steady this analysis yields $\alpha_{steady}=0.28$ and $\beta_{steady}=0.57$, which is consistent with the classical steady heating values given above \citep{1978ApJ...220..643R}. The default EBTEL-impulsive table yields $\alpha_{imp}=0.34\pm0.07$ and $\beta_{imp}=0.44\pm0.04$, which is measurably different from the steady heating case.
Substituting a given parametric heating law described by Eq.\,(\ref{eq:QvsBL}) into Eq.\,(\ref{eq:TvsQL}) we obtain:

\begin{equation}\label{eq:TvsBL}
T \propto  B^{a\alpha} L^{\beta-b\alpha}.
\end{equation}


\begin{deluxetable}{llll} \tablecaption{Geometrical and magnetic field properties for the flux tubes shown in Figure\,\ref{fig_overview}a, averaged over 400 field lines. 
}
\tablehead{\colhead{Parameter}& \colhead{Fluxtube 1}&  \colhead{Fluxtube 2}&  \colhead{Fluxtube 3}}
\startdata
$L\left(10^9cm\right)$& $  22.28\pm   0.21$& $  12.96\pm   0.23$& $   2.04\pm   0.19$\\
$\langle B_{avg}\rangle\left(G\right)$& $    182\pm      3$& $    197\pm      5$& $   1107\pm     59$\\
$\langle B_{-}\rangle_0\left(G\right)$& $    714\pm    156$& $    146\pm     29$& $   1799\pm    206$\\
$\langle B_{-}\rangle_{TR}\left(G\right)$& $    541\pm     61$& $    128\pm     15$& $   1354\pm    273$\\
$\langle B_{top}\rangle\left(G\right)$& $     20\pm      0$& $     32\pm      1$& $    945\pm    135$\\
$\langle B_{+}\rangle_{TR}\left(G\right)$& $   2840\pm    141$& $   2532\pm    122$& $   1082\pm    205$\\
$\langle B_{+}\rangle_0\left(G\right)$& $   3410\pm     13$& $   2984\pm     30$& $   1199\pm    223$\\
\enddata
\tablenotetext{}{
}
\end{deluxetable}
\label{BLtable}

Let us demonstrate that the distribution of the coronal temperature over strong magnetic field regions in the selected AR is highly sensitive to the parametric heating indices. To do so, we employ properties of the magnetic field lines associated with three strong-magnetic-field regions in Figure\,\ref{fig_overview}a. 
{Table\,\ref{BLtable} shows
two metrics needed for our heating model (two top lines) along with
several other metrics, because some models may employ them \citep[see, e.g.,][]{2004ApJ...615..512S}: $L$: magnetic field line length; $\langle{B_{avg}}\rangle$: mean magnetic field along the field line; $\langle B_{-}\rangle_0$ \& $\langle B_{-}\rangle_{TR}$: negative polarity footpoint mean absolute magnetic field at the photospheric and, respectively, transition region heights; $\langle B_{+}\rangle_0$ \& $\langle B_{+}\rangle_{TR}$: positive polarity footpoint mean absolute magnetic field at the photospheric and, respectively, transition region heights; $B_{top}$: mean magnetic field at the loop top. The metrics needed to apply our heating model for the three selected flux tubes are:}
$B_{avg,1}=182$\,G, $L_1=220$\,Mm; $B_{avg,2}=197$\,G, $L_2=130$\,Mm; and $B_{avg,3}=1100$\,G, $L_3=20$\,Mm; see Table\,\ref{BLtable}.
Here, for illustration only, we adopt the impulsive heating indices
$\alpha=\alpha_{imp}=0.34$, $\beta=\beta_{imp}=0.44$, $b=0.75$ (see Table\,\ref{Table_metrix}), and consider two values of $a$: $a=1$ and $a=2$.

Normalizing the volumetric heating rate such as $T_2=2$\,MK, we obtain comparable values of $T_1\approx 2.15$\,MK and $T_3\approx 2.55$\,MK for $a=1$. In contrast, for $a=2$, we obtain remarkably different temperatures of $T_1\approx 2.1$\,MK and $T_3\approx 4.6$\,MK. This estimate implies that the brightness ratios between the radio sources associated with these three different areas will  differ measurably depending on the involved heating mechanism. This sensitivity permits us to substantially constrain the parameters of the coronal heating law.


\subsection{Computing Synthetic Radio Maps}

GX Simulator permits us to  select an arbitrary FOV within the model with arbitrary size of the image pixel. Once selected, the tool solves for intersections between each LOS and the model voxels over a non-uniform grid and creates the LOS information to be used by the radiation transfer codes. This LOS information is then sent to a computing engine (a dynamic link library, dll) that numerically solves the radiative transfer equation \citep{Fleishman2014a}. This radiative transfer includes GR and free-free emission and absorption and also accounts for the polarization transformation along LOS; in particular, when the radiation propagates through quasi-transverse (QT) magnetic field layers, where the mode coupling can be either strong or weak depending of the frequency and ambient plasma parameters. The results of this radiation transfer for each pixel over a selected range of frequencies are put together to form a set of multi-frequency images. The image resolution is defined by the user-selected size of the image pixel, while does not depend on any instrumental resolution.

In our analysis, in all cases, we selected the same $200\arcsec\times200\arcsec$ FOV for all models with the $2\arcsec\times2\arcsec$ pixel size. We made computations over 100 frequencies between 1 and 42\,GHz that include both 17 and 5.7\,GHz for direct comparison with the NoRH and SSRT data.

\subsection{PSF-convolution of synthetic maps and their  co-alignment with the observed ones}
\label{S_fine_tune}


To compare quantitatively the synthetic and observed radio maps, we performed the following steps:

a) The synthetic radio map was convolved with the instrument PSF and shifted to correct for the possible instrumental positioning inaccuracy (see below). We used the model PSFs of NoRH (provided by the SolarSoft NoRH package) and SSRT (using the code provided by V. Grechnev, private communication) corresponding to the times of observations. The model NoRH beam was corrected (degraded) by a factor of 1.185 to improve the overall comparison with the observed images; this is similar to a `Clean Beam Width Factor' often applied while generating \rhessi\ images \citep{2010ApJ...717..250K}.

b) The observed radio map was interpolated and truncated to match the resolution and FOV of the synthetic map, respectively, which enabled pixel-by-pixel comparison.

Since interferometric instruments do not always provide an accurate image positioning, we applied an additional 2D shift to the synthetic radio maps\footnote{Technically, this is more convenient than to shift the observed maps.} to provide a better alignment with the observed ones. The best alignment was achieved by maximizing the pixel-by-pixel cross-correlation coefficient $r(\Delta x, \Delta y)$ of two radio maps.
We searched for the local maximum of the cross-correlation coefficient using the steepest-descent method, starting with zero shift. In all cases, the optimum shift proved to be smaller than the instrument PSF width. We computed the optimum shift using the intensity maps, and the same shift was then applied to the polarization maps.
After this co-alignment, a cut-out of the observational map with the same $200\arcsec\times200\arcsec$ FOV (as in the synthetic map) was used in the analysis.

\begin{figure}
\centering
\includegraphics[width=0.98\linewidth, bb=90 100 500 688]{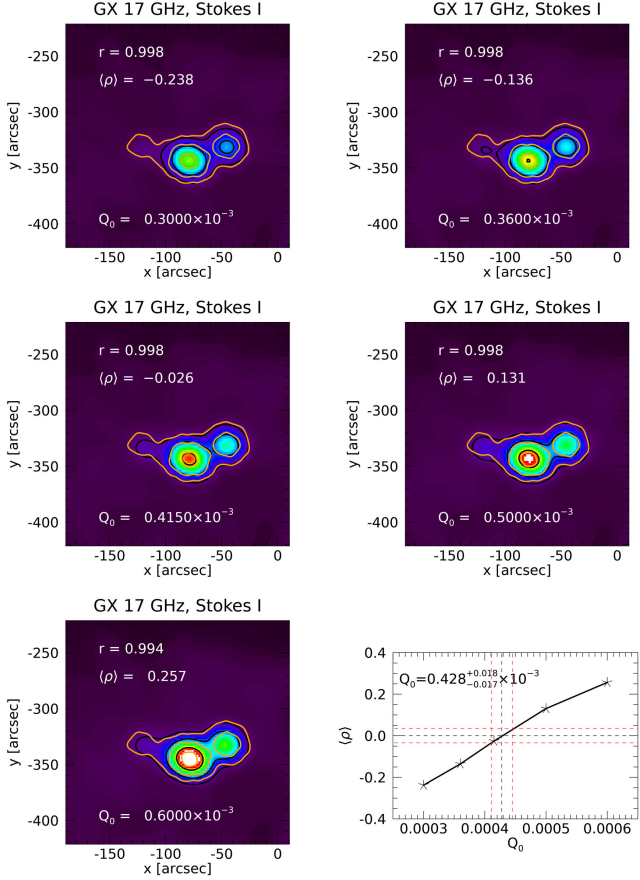}
\caption{Illustration of the procedure used to find the best heating rate for  a given pair of $a$ and $b$ exponents in the parametric heating law. Five synthetic images are computed for different heating rates $Q_0$ and displayed as background maps and [12; 30; 80]\% black contour levels, to be compared with the observed NoRH 17 GHz image, overlaid as [12; 30; 80]\% orange contour levels. 
The computed Pearson correlation coefficients $r$ and the $\left<\rho\right>$ metrics are computed and displayed in each case. The $\left<\rho\right>$ metrics are plotted in the last panel as a function of $Q_0$, the data are interpolated, and the intersection with the level zero (black dashed lines) gives the best $Q_0$ for the given pair of $a$ and $b$. Intersections with the edges of the $\left<\rho\right>$ confidence interval (red dashed lines) provide the $Q_0$ confidence interval.  \label{Fig_ftune}}
\end{figure}

\subsection{Selecting best heating rate for a given parametric heating model\\ 
}
\label{S_ftune}
Once we have selected a coronal magnetic field structure, the synthetic radio emission map becomes only dependent on the average heating rate $Q_0$ and the heating model parameters $a$ and $b$; thus our aim is to find the combination of these three parameters that provides the best match to the observations. For a given combination $(a, b)$, we searched for the best-match heating rate $Q_0$ using the total emission intensity (which increases as the heating rate increases). The total synthetic radio emission intensity (in sfu) from the considered active region  $I_{\mathrm{total}}^{\mathrm{mod}}(Q_0)$ was computed by integrating the emission intensity over the region of interest (ROI) above a given brightness threshold (in this case adopted to be 12\% of the map brightness peak), for several representative values of the heating rate $Q_0$ (see Fig. \ref{Fig_ftune}). Then the dependence of $\log I_{\mathrm{total}}^{\mathrm{mod}}(Q_0)$ on $\log Q_0$ was interpolated by a third-degree polynomial. The best-fit value of $Q_0$ was found by solving the equation $I_{\mathrm{total}}^{\mathrm{mod}}(Q_0)=I_{\mathrm{total}}^{\mathrm{obs}}$, where $I_{\mathrm{total}}^{\mathrm{obs}}$ is the total observed radio emission intensity from a ROI within a 12\%  contour of the observed radio map obtained as described in Section \ref{S_fine_tune}. The confidence range of $Q_0$ can be estimated from the relation $I_{\mathrm{total}}^{\mathrm{mod}}(Q_0)=I_{\mathrm{total}}^{\mathrm{obs}}\pm\Delta I_{\mathrm{total}}^{\mathrm{obs}}$, where $\Delta I_{\mathrm{total}}^{\mathrm{obs}}$ is the instrumental uncertainty in determining the absolute radio flux. This algorithm can be implemented by equating to zero one of the following metrics:

\begin{equation}
\label{Eq_res2_ideal_S2}
\left<{\cal R}\right>=\frac{1}{N}
\sum_{i=1}^{N}
\left( {x_i}-{y_i}\right),
$$$$
\left<\rho\right>=\frac{1}{N}\sum_{i=1}^{N} \left( \frac{x_i}{y_i}-1\right)    ,
\end{equation}
where  $x_i$ and $y_i$ are the model and, respectively, the observed brightness in pixel $i$, and $N$ is the total number of valid pixels for which the comparison is made.

\subsection{2D Cross-Correlations, Residuals, and Other Metrics of Success }
\label{S_metrics}


To quantify goodness of models we want to use normalized sum of squares of the individual residuals:
\begin{equation}
\label{Eq_res2_ideal}
\left<\rho^2\right>=\frac{1}{N}\sum_{i=1}^{N}
\left( \frac{x_i}{y_i}-1\right)^2    ,
\end{equation}
or

\begin{equation}
\label{Eq_res2_ideal_S}
\left<{\cal R}^2\right>=\frac{1}{N}\sum_{i=1}^{N}
\left( {x_i}-{y_i}\right)^2.
\end{equation}

However, these measures are biased for the following reasons.
In Section\,\ref{S_fine_tune} we used a residual $\left<\rho\right>$ computed within a user-defined area of the image to specify the best heating rate $Q_0$ for each given case. Ideally, we must have $\left<\rho\right>=0$ and $\left<{\cal R}\right>=0$, but in reality they are not zeros. A more important is that these (best) residuals are different for various models. This means that each `best $Q_0$' synthetic image is systematically off by its unique (small) value compared with the observed image.

To make up for such potentially non-zero residuals, we adopt following metrics of success:
\begin{equation}
\label{Eq_res2_0}
\sigma_\rho^2=\frac{1}{N}\sum_{i=1}^{N}
\left( \frac{x_i}{y_i}-1-\left<\rho\right>\right)^2 = \left<\rho^2\right>-\left<\rho\right>^2 
$$$$
\sigma_{\cal R}^2=\frac{1}{N}\sum_{i=1}^{N}
\left( {x_i}-{y_i}- \left<{\cal R}\right>\right)^2=
\left<{\cal R}^2\right> - \left<{\cal R}\right>^2,
\end{equation}
which reduce to Eq.\,(\ref{Eq_res2_ideal}) and  (\ref{Eq_res2_ideal_S}) in the ideal case of zero averaged residuals.

Original image observed by NoRH contains two well separated microwave sources associated with two umbras of a complex sunspot located in the middle of AR 11520.
To quantify the ability of different models to reproduce the relative locations of these sources, we introduced an additional metric $d_{mod}/d_{obs}$, where $d_{mod}$ is the distance between the brightness peaks of these two sources in a synthetic map, while $d_{obs}$ is the distance that was actually observed by NoRH at 17 GHz.


	
\subsection{Validation and selection of the best magneto-thermal model
}

In this study, we employ distribution of the brightness temperature of Stokes I, $T_{I,17}$, at 17\,GHz to fine tune the heating rates $Q_0$ and to then select the best $a$ and $b$ pairs for each 3D magnetic model. This forms a set  of  best thermal models, whose metrics of success (Table\,\ref{Table_metrix}) can be compared to each other in order to select the very best model.

\begin{figure}
\centering
\includegraphics[width=0.98\linewidth]{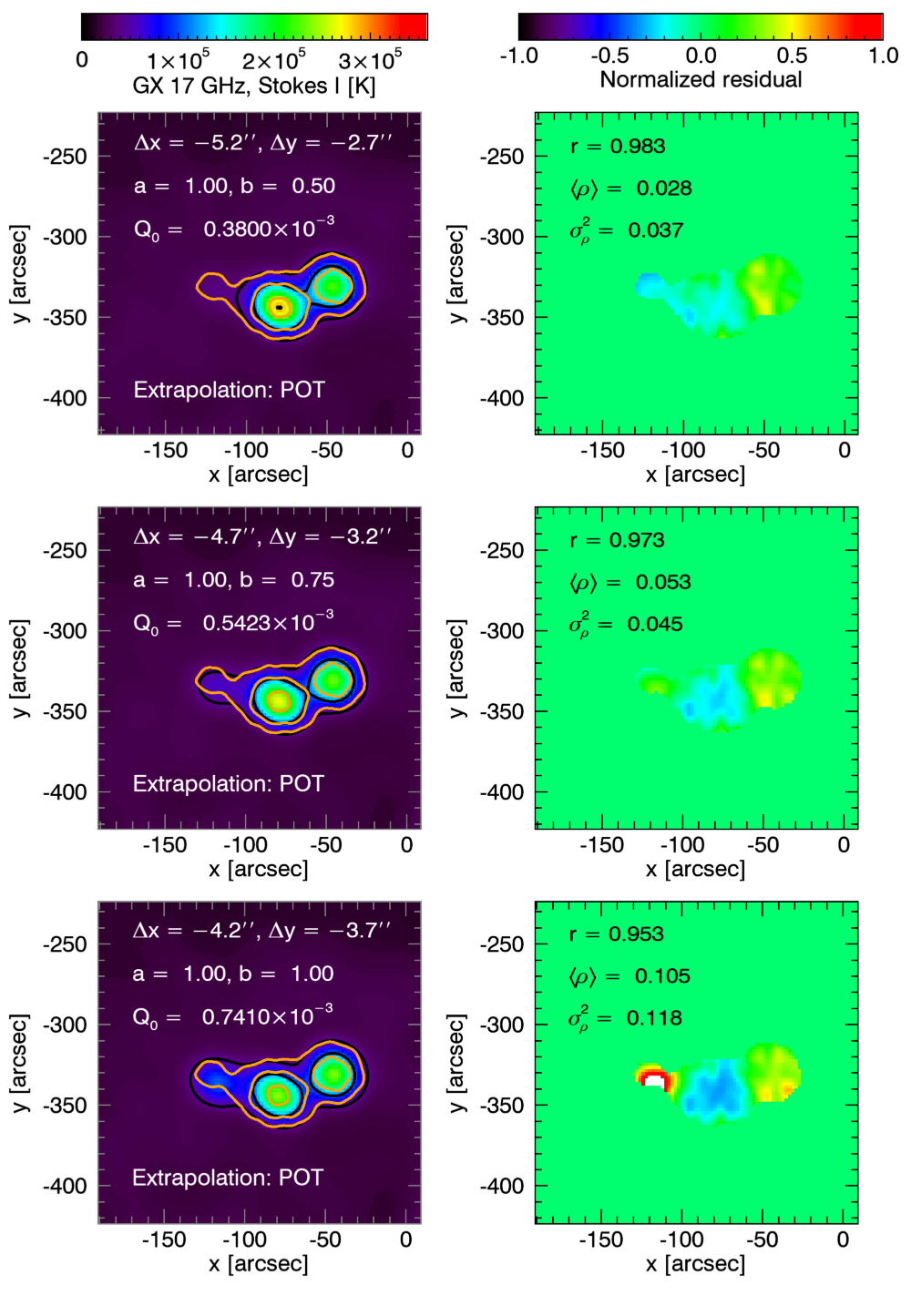}
\caption{Synthetic and residual maps for thermal models produced using POT extrapolation. Left column: PSF-convolved best-$Q_0$ synthetic maps obtained for various parametric heating laws (indices $a$ and $b$ are printed in the panels) are shown as background images and black contours at [12, 30 and 80]\% of the maximum brightness temperature.  
The shift ($\Delta x$ and $\Delta y$) needed to co-align the model image with the observed one is printed in the panels. The observed 17\,GHz NoRH images are shown in yellow at the same contour levels. {The color scale in each column is shown by the colorbars placed at the top of the Figure.} Right column: the associated residual maps. The cross-correlation coefficient ($r$), mean residual ($\left<\rho\right>$) and variance ($\sigma_\rho^2$) are printed in the panels.
}
\label{Fig_POT_BestQ0}
\end{figure}

In addition, we use two more metrics to validate the magneto-thermal models. One of them is the brightness distribution of the Stokes V parameter, $T_{V,17}$, at 17\,GHz. This gives a complementary information to the $T_{I,17}$ distribution because the absolute values of both $T_I$ and $T_V$ depend on local conditions at the gyro layer, while the sign of $T_V$ depends also on conditions along the LOS: if there is (are) QT layer(s) along the LOS, the sense of polarization can be affected. Thus, having a correct $T_{V,17}$ distribution additionally justifies the magnetic model at high heights of the data cube. This is important because building the magnetic field lines, used to create the thermal model, depends on the magnetic field at those high heights. Another measure sensitive to the magnetic field at high heights is the radio brightness at a low frequency, which forms high in the corona. Here we employ the radio brightness at the available 5.7\,GHz SSRT frequency.

\begin{figure}
\centering
\includegraphics[width=0.98\linewidth]{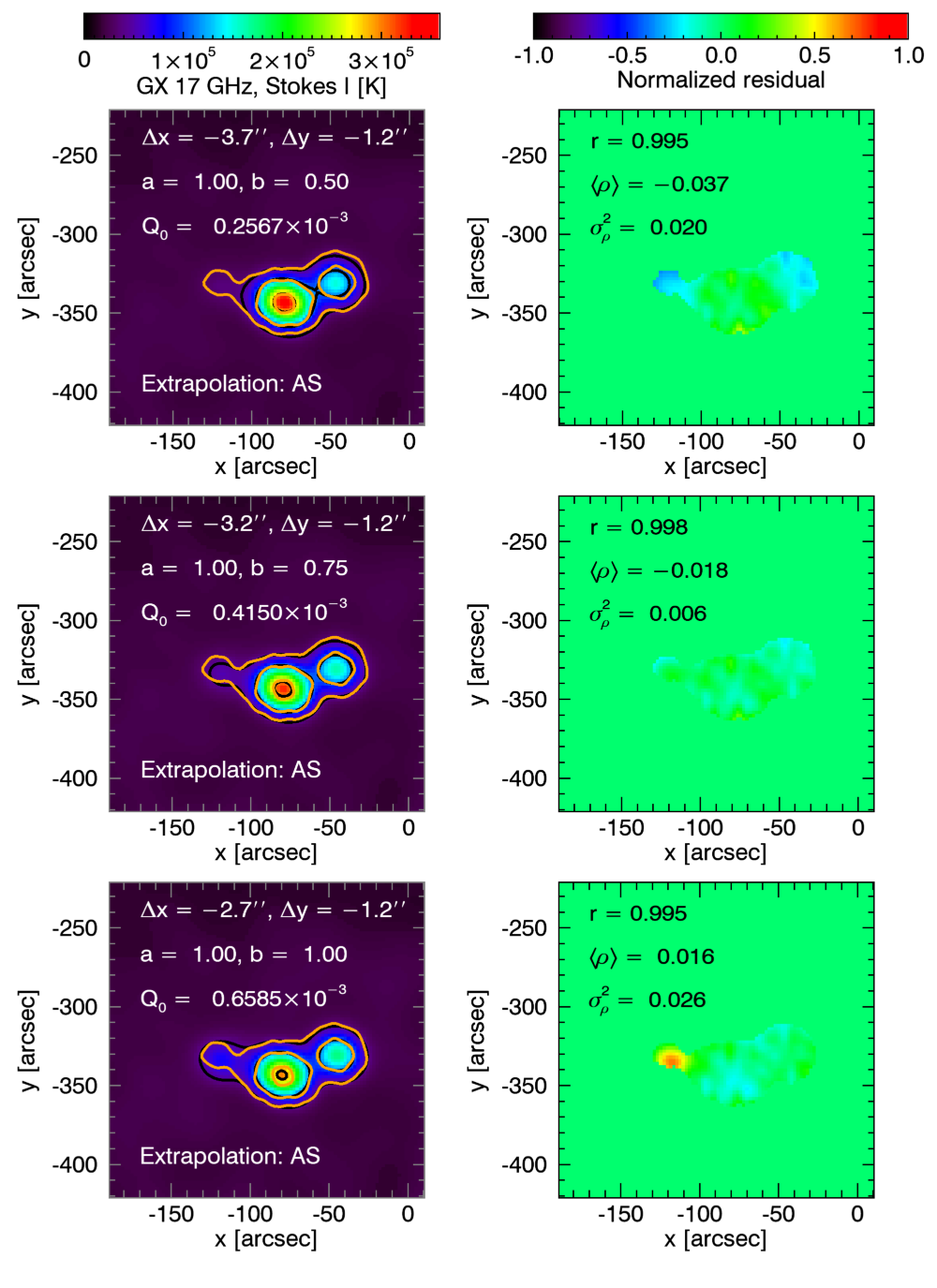}
\caption{Same as in Figure\,\ref{Fig_POT_BestQ0} but for AS NLFFF extrapolation. The color scale in each column is shown by the colorbars placed at the top of the Figure.} 
\label{Fig_AS_BestQ0}
\end{figure}	

\newpage	
\section{Magneto-thermal models, their comparison and validation}

To create the magneto-thermal models, we used four different magnetic models---one POT and three NLFFF extrapolations, see Section\,\ref{S_POT_NLFFF}, using exactly the same FOV, model sizes, and spatial resolution. To select the best parametric coronal heating model (the best combination of $Q_0$, $a$, and $b$) we performed a number of initial tests, from which we concluded that models with $a>1$ do not offer any good morphological match between synthetic and observed $T_{I,17}$ images: they all predicted measurably stronger emission from the eastern footpoint of Flux Tube\,3 than observed. 
\citet{2019ApJ...877..129U} found that $a$ and $b$ are not independent, so that an increase of $a$ can be (partly) compensated by a decrease of $b$ and \textit{vice versa}.
Therefore, we adopted a single $a=1$ value 
and considered three different values of $b=1,\,0.75,\,0.5$. For each $a$ and $b$ pair, we identified the best $Q_0$ as described in Section\,\ref{S_ftune}
; Figure\,\ref{Fig_ftune}. A more systematic study of the $a-b$ grid is underway.

To compare the PSF-convolved best-$Q_0$ synthetic maps  with the observed $T_{I,17}$ maps, we co-align them as described in Section\,\ref{S_fine_tune}, compute and plot the residual maps, and compute metrics of success as described in Section\,\ref{S_metrics}. This set of maps for the reference case of POT extrapolation is shown in Figure\,\ref{Fig_POT_BestQ0}, while the best NLFFF case obtained with the AS code is given in Figure\,\ref{Fig_AS_BestQ0}; we do not show maps for two other NLFFF extrapolations, IM and IM0, as the AS NLFFF extrapolation outperforms the other ones  for this instance of AR 11520; see below.   Metrics of success of these `best $Q_0$' models are given in Table\,\ref{Table_metrix}. The best metrics 
for each magnetic cube are highlighted by the boldface in Table\,\ref{Table_metrix}. 
Note that for various magnetic models, their best heating laws are not identical to each other. For example, we  found $b=0.5$ to be the best for the POT model,  and $b=0.75$ for  the AS model. 

Between four best  magneto-thermal models highlighted in Table\,\ref{Table_metrix} we now  select the very best one by comparing their $\sigma_\rho^2$ and other metrics shown in Table\,\ref{Table_metrix}. This favors the AS model with $a=1$ and $b=0.75$. This choice, made based on the numeric metrics, is consistent with  the visual  inspection of the residual maps: the middle row residual map in Figure\,\ref{Fig_AS_BestQ0} is almost uniformly green (near zero residuals); thus, the model reproduces the data equally well throughout the entire ROI.

\begin{figure}
\centering
\includegraphics[width=0.98\linewidth]{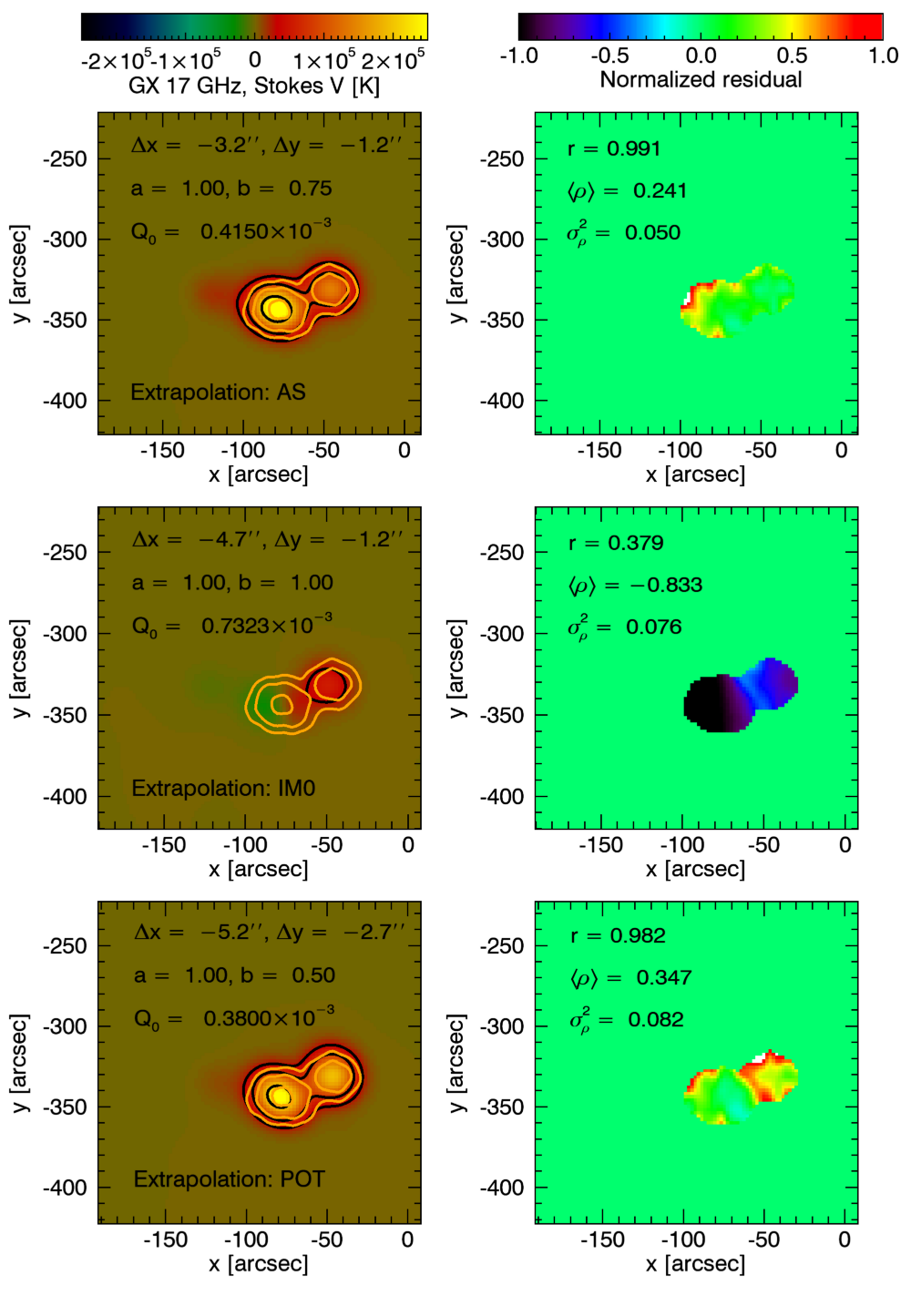}
\caption{Comparison between synthetic and observed Stokes\,V for three best models: AS, IM0, and POT. The color scale in each column is shown by the colorbars placed at the top of the Figure. The panel layout is the same as in Figure\,\ref{Fig_POT_BestQ0}. }
\label{Fig_TV_BestQ0}
\end{figure}

\begin{figure}
\centering
\includegraphics[width=0.98\linewidth]{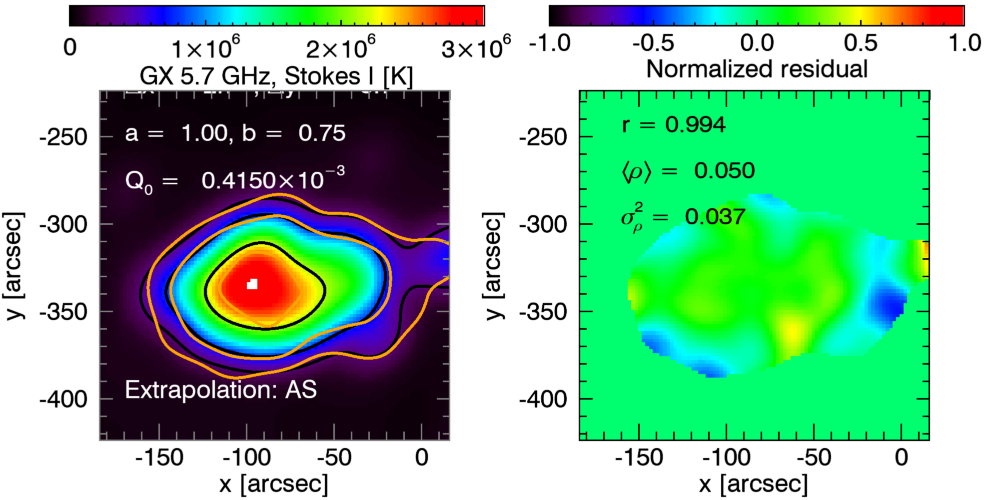}
\caption{Models-to-data comparison for AS extrapolation at 5.7\,GHz assuming the optimal indices $a=1$ and $b=0.75$ according to the best case in Table~\ref{Table_metrix}. Contours correspond to  [12, 30 and 80]\% of the maximum brightness temperature. 
}
\label{Fig_AS_a1b075_5.7GHz}
\end{figure}

\begin{figure}
    \centering
    \includegraphics[width=0.95\linewidth]{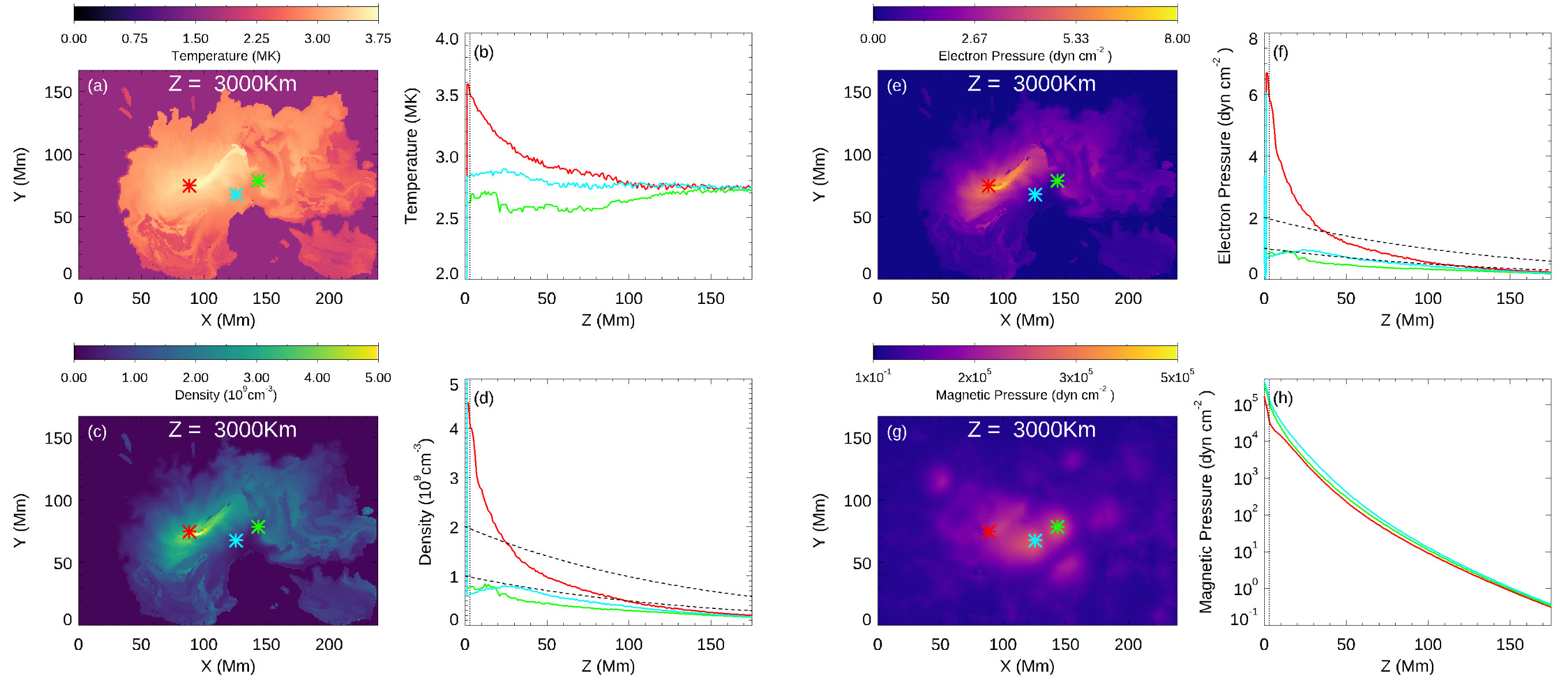} 
    \caption{Electron temperature (a,b),  density (c,d), and pressure (e,f) distributions in the model box. For the reference, panels (g,h) show distribution of the magnetic pressure.  Images show horizontal distributions of the corresponding values at $Z=3$\,Mm. Plots show vertical dependence above three locations marked by colored symbols in the associated images. Panels (d,f) also show two hydrostatic solutions for $T=3$\,MK and two different initial values at $Z=0$. We also provide an animated version of this figure that, using the same layout, displays the density and temperature distributions as the vertical coordinate changes from the bottom to the top of the model box. The vertical coordinate is plotted in the images; location of the displayed plane is indicated by moving vertical line in panels (b,d,f,h). The animation length is 8 seconds.}
    \label{fig:nT_vs_height}
\end{figure}

Additional confirmation in favor of this selected model comes from  the inspection of other metrics given in Table\,\ref{Table_additional_metrix}. In particular, the correlation coefficient between the synthetic and observed polarization maps, $T_{V,17}$, is 99.1\%  (Figure\,\ref{Fig_TV_BestQ0}), while the corresponding residual map is almost uniform, with $\sigma_\rho^2=0.05$. The brightness temperature distribution at 5.7\,GHz is also reproduced well, see Figure\,\ref{Fig_AS_a1b075_5.7GHz}, in particular, the brightness peak projects on the neutral line as observed  {because the hottest plasma is located there as illustrated in Figure\,\ref{fig:nT_vs_height}}; the correlation coefficient is 99.4\% and $\sigma_\rho^2=0.037$. 
We do not put too much quantitative weight on this $T_{I,5.7}$ metrics because the SSRT image is obtained using a long integration; thus, both magnetic and thermal structures could evolve during this time. For the same reason, we do not use the polarization map at 5.7\,GHz. We conclude that the AS model with $a=1$ and $b=0.75$ is our best magneto-thermal model, which truthfully reproduces both the magnetic and thermal structures of AR 11520 on 2012-Jul-12 05:00\,UT. The thermal structure of this best model is illustrated in Figure\,\ref{fig:nT_vs_height}. {This animated Figure shows the 3D distribution of electron temperature, density, and pressure along with the magnetic field pressure. Importantly, despite of a relatively simple heating law that depends on only three parameters, the thermal structure shows a lot of complexity due to the underlying complexity of the magnetic skeleton above the 1000\,km height. Below that height, only the standard chromospheric model described in \citet{Nita2018} is present, and thus not shown, while between 1000 and 2000\,km there are both coronal and chromospheric voxels depending on the underlying chromospheric structure. Panels (d,f), for comparison, show the hydrostatic atmosphere solutions (dashed curves) for $T=3$\,MK. Note that at high heights the numerical solutions are rather close to the hydrostatic ones, while they deviate from each other strongly in the low corona; especially, for the red line that corresponds to the flux tube associated with the neutral line. This is because, in the magnetically closed corona, the thermal structure primarily depends on conditions along the magnetic field lines, rather than on the height.}



\begin{deluxetable*}{lcccccccc}
\tablecaption{Metrics of the  magneto-thermal models agreement with Stokes I images obtained at 17 GHz. All metrics are computed within the area where the observed intensity at 17 GHz is above 12\,\% of its maximum value. The last column shows the ratio of distances between radio sources in a synthetic image and observations $d_{mod}/d_{obs}$. The best metrics for each magnetic model are highlighted with the bold font. 
}
\tablehead{
\colhead{Model}&
\colhead{a}&
\colhead{b}&
\colhead{r}&
\colhead{$\left<{\cal R} \right>$}&
\colhead{$\sigma^2_{\cal R} \times 10^{-9}$}&
\colhead{$\left<\rho\right>$}&
\colhead{$\sigma^2_\rho$}&
\colhead{$d_{mod}/d_{obs}$,}
}
\startdata
AS& 1.00&0.50&0.995&-2073.& 0.150&-0.037& 0.020& 0.97\\
AS& 1.00&0.75&\textbf{0.998}&-3035.& \textbf{0.079}&-0.018& \textbf{0.006}& \textbf{1.01}\\
AS& 1.00&1.00&0.995&-3945.& 0.232& 0.016& 0.026& 0.57\\
IM& 1.00&0.50&\textbf{0.974}& -586.& \textbf{0.839}& 0.003& \textbf{0.082}& 0.87\\
IM& 1.00&0.75&0.963&  261.& 1.182& 0.034& 0.093& 0.87\\
IM& 1.00&1.00&0.951& -386.& 1.544& 0.062& 0.091& 0.87\\
IM0& 1.00&0.50&0.989& -457.& 0.376&-0.043& 0.048& 0.92\\
IM0& 1.00&0.75&0.993& -170.& 0.234&-0.026& 0.032& 0.94\\
IM0& 1.00&1.00&\textbf{0.995}&  -36.& 0.149&-0.006& \textbf{0.016}& \textbf{0.96}\\
POT& 1.00&0.50&\textbf{0.983}& -169.& \textbf{0.560}& 0.028& \textbf{0.037}& 0.98\\
POT& 1.00&0.75&0.973&-1133.& 0.896& 0.053& 0.045& \textbf{1.01}\\
POT& 1.00&1.00&0.953&-1700.& 1.526& 0.105& 0.118& 1.08\\

\enddata
\end{deluxetable*}
\label{Table_metrix}

\begin{deluxetable*}{lcc|ccc|ccccc}
\tablecaption{Metrics of the  magneto-thermal models agreement with Stokes V images obtained at 17 GHz and Stokes I images at 5.7 GHz for the full field of view. 
}
\tablehead{
\colhead{}&
\colhead{}&
\colhead{}\vline&
\multicolumn{3}{c}{17\,GHz (V)}\vline&
\multicolumn{5}{c}{5.7\,GHz (I)}\\
\colhead{Model}&
\colhead{$a$}&
\colhead{$b$}\vline&
\colhead{$r$} &
\colhead{$\left<{\cal R}\right>$}&
\colhead{$\sigma^2_{{\cal R}}\times 10^{-9}$}\vline&
\colhead{$r$}&
\colhead{$\left<{\cal R}\right>$}&
\colhead{$\sigma^2_{{\cal R}}\times 10^{-9}$}&
\colhead{$\left<\rho\right>$}&
\colhead{$\sigma^2_\rho$}
}
\startdata
AS& 1.00& 0.75&\textbf{0.989}&  3529& \textbf{0.087}&\textbf{0.992}& 45341&\textbf{16.378}& 0.429& \textbf{0.511}\\
IM0& 1.00& 1.00&0.372&-11727& 1.085&0.990& 75853&30.332& 0.496& 0.566\\
POT& 1.00& 0.50&0.981&  4404& 0.121&0.978& 65931&27.131& 0.688& 1.160\\

\enddata
\end{deluxetable*}
\label{Table_additional_metrix}

\section{Discussion and conclusions}


Starting from \citet{1977ApJ...213..278K}, spatially resolved \mw\ observations from many instruments, such as VLA, RATAN-600, OVSA, NoRH, and SSRT, have been employed to infer both magnetic and thermal properties of ARs \citep[see, e.g.,][and references therein]{1980A&A....82...30A,1994ApJ...420..903G,2010ARep...54...69T,2010AstBu..65...60K,2011ApJ...728....1T,2012ARep...56..790K,2015ApJ...805...93W,2018SoPh..293...13S,2019SoPh..294...23A}. These and other studies reported distributions of radio brightness temperature indicative of coronal temperature distribution over the corresponding gyroresonant surfaces. These temperature distributions were then put in a framework of a model to constrain coronal parameters. While some studies \citep{1979AZh....56..562G,1985A&A...143...72K,2010AstBu..65...60K} utilized simplified (e.g., dipole) magnetic models, others used more realistic potential or force-free extrapolations \citep{1980A&A....82...30A,2011ApJ...728....1T,Nita_etal_2011,2018SoPh..293...13S,2019SoPh..294...23A}. The associated thermal models relied on various assumptions, such as a plane-parallel stratification of the corona, monotonic behavior of coronal parameters along a LOS, hydrostatic equilibrium or others, which may or may not be valid. More importantly is that those thermal models were decoupled from the underlying magnetic skeletons.

The coupling between the thermal and magnetic structures is crucial for understanding coronal heating because it is the magnetic energy that is somehow dissipated to support the hot corona. This dissipated magnetic energy is deposited on individual magnetic flux tubes; thus, thermal properties of these flux tubes are the outcome of the field-aligned (1D) hydrodynamics \citep{Klimchuk2006,Klimchuk2015}. Our novel approach to the \mw\ modeling closely packs the 3D volume with those individually heated flux tubes such as each model voxel associated with a closed magnetic field line is being populated by a realistic thermal plasma within   the GX Simulator methodology \citep{Nita2018}. Having the thermal plasma defined in all voxels of the model, rather than along a subset of bright loops, makes the magneto-thermal model realistic, as shown in Figure\,\ref{fig:nT_vs_height}. This approach has been successfully applied to modeling \mw\ emission from AR 12673, which had a record-breaking coronal magnetic field \citep{Anfinogentov2019}.
However, \citet{Anfinogentov2019} have not specifically considered the coronal heating problem.

AR 11520, which we model here, is perfectly suited for addressing the coronal heating problem, because its GR emission at 17\,GHz comes from three different sources with remarkably different connectivities; see Figure\,\ref{fig_overview}a 
and Table\,\ref{BLtable}. This means that various parametric heating laws described by Eq.\,(\ref{eq:QvsBL}) with different $a$ and $b$ will result in measurably distinct distributions of the synthetic GR brightness. We emphasize that dependence on both $B_{avg}$ and $L$ in Eq.\,(\ref{eq:QvsBL}) is needed. Even though, there is a general trend that $B_{avg}$ is getting smaller for larger $L$, our attempt to find a linear regression between these two values has failed. Similarly to  \citet{Mandrini2000}, we found that a nonlinear relationship links $B_{avg}$ and $L$ much better than a linear regression; see Figure\,\ref{fig:BQL}.

It is further important that the individual magnetic field lines, which control the parametric heating properties, are computed from the entire magnetic model. Therefore, even the GR emission, formed very low in the corona, is sensitive to what happens at higher heights of the magnetic skeleton. Thus, if we have obtained the synthetic brightness distribution matching observations, this match validates both magnetic and thermal models of the AR at once. This permits  solid conclusions on the preferred parametric heating law based on the data-validated 3D model.
Here we have employed these unique properties of the GR emission to select the best heating model, which has  $a=1$ and $b=0.75$.

The GX Simulator methodology has been applied by \citet{2019AGUFMSH41F3323S} to  modeling EUV emission from ARs.
\citet{2019AGUFMSH41F3323S} have 
tested three favorite models: (i) `critical angle' with $a=2$ and $b=1$, (ii) `critical twist' with $a=2$ and $b=2$, and (iii) `resonance' with $a=0$ and $b=2$. They rejected all three models, which is consistent with our best model that has $a=1$ and $b=0.75$.

\begin{figure}
\centering
\includegraphics[width=0.98\linewidth]{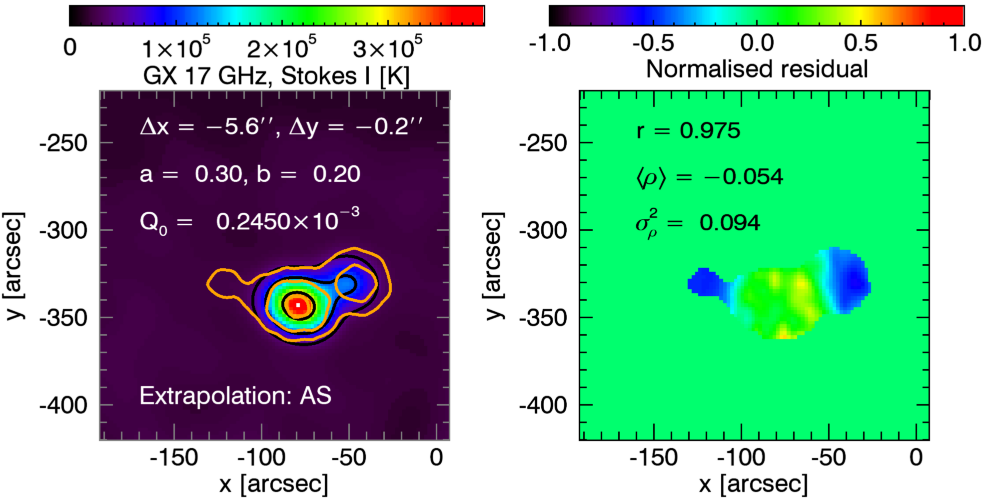}
\caption{Models-to-data comparison for AS NLFFF extrapolation at 17\,GHz assuming the optimal indices $a=0.3$ and $b=0.2$ reported by \citet{2019ApJ...877..129U}. Contours correspond to  [12, 30 and 80]\% of the maximum brightness temperature.}
\label{Fig_AS_a03b02}
\end{figure}

\citet{2019ApJ...877..129U} performed a systematic search over the $a$ and $b$ parameter space {in another AR} by isolating and analyzing individual bright loops in EUV and associated magnetic flux tubes in NLFFF reconstructions of ARs. They concluded that the best heating model has  $a=0.3$ and $b=0.2$ with the range of uncertainties $\sim\pm 0.2$. They noted that $B_{avg}$ and $L$ are correlated to each other; thus, dependencies of them are not independent---larger $a$ requires smaller $b$ and vice versa. To directly test if their model could be consistent with \mw\ emission from AR 11520, we adopted $a=0.3$ and $b=0.2$, determined best $Q_0$ as described in Section\,\ref{S_fine_tune}, and computed synthetic maps and their residuals with the data. Figure\,\ref{Fig_AS_a03b02} explicitly shows that the GR brightness distribution computed from this model does not agree with the data; thus, it can firmly be rejected in the case of AR 11520.

There could be both methodological and physical reasons for this apparent disagreement. In our study we explicitly demonstrated high sensitivity of the modeling results to the employed magnetic model. In contrast to the \mw\  case, the main problem of the model validation with EUV emission is its insensitivity to the magnetic field, aside from the shape of the bright EUV loops. This means that an important step of quantitative validation of the magnetic model by data is missing in the EUV case. A possible physical reason is that we modeled an averaged thermal structure in the AR volume, while \citet{2019ApJ...877..129U} considered individual bright loops. It is possible that average heating and enhanced ``selective'' heating that supports bright loops standing out cleanly against the background are different from each other. Clearly, this question deserves further analysis using more data and more ARs.

The fact that a 3D magneto-thermal model, built using a parametric heating law {that depends on only three free parameters}, matches observations very closely is a great success of the heating parametrization in the form of Eq.\,(\ref{eq:QvsBL}).
{We emphasize that our thermal model pertains to a smooth (i.e., average) coronal component, rather than bright coronal loops. The use of the NoRH microwave images with a moderate spatial resolution is well suited to constrain this average component because the radio brightness is itself naturally averaged due to this moderate resolution. It is likely that observations with higher spatial resolution will favor a more sophisticated model with $Q_0$, $a$, and $b$ dependent on the spatial location.  }

It would now be interesting to link our best model  ($a=1$ and $b=0.75$) with a specific heating mechanism. \citet{Mandrini2000} reviewed a range of the proposed coronal heating mechanisms and listed corresponding parametrizations in  their Table\,5, which, however, does not include any case with $a=1$ and $b=0.75$. Perhaps, their case closest to ours is $a=1$ and $b=1$, which is clearly not favored by our modeling; see the bottom panel in Figure\,\ref{Fig_AS_BestQ0} and our Table\,\ref{Table_metrix}. The scaling laws listed by \citet{Mandrini2000} depend not only on $B_{avg}$ and $L$, but also on other parameters such as plasma density and transverse velocity at the base of the corona. Dependences of the heating rate on these parameters could be replaced by some dependence on $B_{avg}$ and/or $L$, if there is a corresponding correlation between parameters \citep{Mandrini2000}. In this case, the indices in the scaling form of Eq.\,(\ref{eq:QvsBL}) may differ from the indices showing explicit dependence on $B_{avg}$ and $L$ in Table\,5 of \citet{Mandrini2000}.

Another remaining unknown is  the dependence of our modeling results on the dynamics of the coronal heating. Here we employed a default EBTEL impulsive heating lookup table used in GX Simulator \citep{Nita2018}, which assumes heating episodes with a triangular shape lasting for 20\,s and repeating every 10,000\,s. Note that this impulsive heating model is different from those adopted by \citet{2019ApJ...877..129U} or \citet{2019AGUFMSH41F3323S}.   Other heating scenarios have to be studied to quantify how sensitive are the modeling results to the dynamics of the impulsive heating.

In this study, we have addressed the coronal heating problem with 3D modeling using only a single frequency (17\,GHz) for the model fine tuning and validation and one more frequency (5.7\,GHz) for the model `sanity' check. Even with this limited set of observational data, we  have obtained very stringent constraints on possible parametric heating law: $a=1$ and $b=0.75$. Uncertainties of these indices can be roughly estimated as $\lesssim\pm0.25$ based on the fact that our best, validated (AS NLFFF), magnetic model produces measurably stronger model-to-data mismatch for either $b=0.5$ or 1  compared with the best thermal  model ($b=0.75$).
We conclude that the use of  the GR \mw\ emission is very promising for addressing the coronal heating problem. A particular progress is expected form multi-wavelength imaging data available from VLA for a number of events \citep[e.g.,][]{2019AGUFMSH41B..05B} and expected soon from solar-dedicated imaging instruments such as EOVSA \citep{2020AAS...23538501G}.

\acknowledgements
This work was partly supported by 
NSF AST-1820613 
and AGS-1743321 
grants
and NASA grants
80NSSC18K0667, 
80NSSC19K0068, 
80NSSC18K1128, 
and 80NSSC20K0627 (SolFER DRIVE center) 
to New Jersey Institute of Technology,
the RFBR research grants 18-32-20165 \verb"mol_a_ved" and 18-29-21016 mk, 
and budgetary funding of Basic Research program II.16. 
The authors are sincerely thankful to Dr. Ivan Mysh'yakov for providing magnetic models and fruitful discussions.

\bibliography{AR11520,fleishman,ms_AR,all_issi_references}
\end{document}